\documentclass[12pt]{article}
\usepackage{latexsym}
\usepackage{amssymb}
\usepackage{amsmath}
\usepackage{graphicx}
\usepackage{slashed}
\usepackage{hyperref}

\def\be{\begin{equation}}
\def\ee{\end{equation}}
\def\bear{\begin{eqnarray}}
\def\eear{\end{eqnarray}}

\newcommand\bra[1]{{\langle {#1}|}}
\newcommand\ket[1]{{|{#1}\rangle}}

\def\o{\omega}
\def\bra{\langle}
\def\ket{\rangle}

\def\o{\omega}

\newcommand{\tn}[1]{\mbox{\tiny #1}}
\renewcommand{\@}[1]{\sqrt{#1}}
\renewcommand{\le}[1]{\label{#1}\end{eqnarray}}
\newcommand{\bea}{\begin{eqnarray}}
\newcommand{\eea}{\end{eqnarray}}

\def\ffract#1#2{\raise .35 em\hbox{$\scriptstyle#1$}\kern-.25em/
\kern-.2em\lower .22 em \hbox{$\scriptstyle#2$}}

\def\half{{1\over2}\,}

\begin{document}
\pagestyle{empty}

\centerline{{\Large \bf Spacetime and Physical Equivalence}}
\vskip1cm

\begin{center}
{\large Sebastian De Haro}\\
\vskip 1truecm
{\it Trinity College, Cambridge, CB2 1TQ, United Kingdom}\footnote{Forthcoming in {\it Beyond Spacetime.
The Foundations of Quantum Gravity,} by N.~Huggett and C.~W\"uthrich (Eds.), Cambridge University Press, 2020.}\\
{\it Department of History and Philosophy of Science, University of Cambridge\\
Free School Lane, Cambridge CB2 3RH, United Kingdom}\\

\vskip .7truecm
{\tt sd696@cam.ac.uk}

\vskip 1cm
\today
\end{center}

\vskip 1cm

\begin{center}

\textbf{\large \bf Abstract}
\end{center}


In this essay I begin to lay out a conceptual scheme for: (i) analysing dualities as cases of theoretical equivalence; (ii) assessing when cases of theoretical equivalence are also cases of physical equivalence. The scheme is applied to gauge-gravity dualities. I expound what I argue to be their contribution to questions about: (iii) the nature of spacetime in quantum gravity;
(iv) broader philosophical and physical discussions of spacetime.

(i)-(ii) proceed by analysing duality through four contrasts. A {\it duality} will be a suitable isomorphism between models: and the four relevant contrasts are as follows: 

(a) {\it Bare theory}: a triple of states, quantities, and dynamics endowed with appropriate structures and symmetries; {\it vs.~interpreted theory}: which is endowed with, in addition, a suitable pair of interpretative maps.

(b)  {\it Extendable vs.~unextendable theories}: which can, respectively cannot, be extended as regards their domains of application.

(c) {\it External vs.~internal interpretations}: which are constructed, respectively, by coupling the theory to another interpreted theory vs.~from within the theory itself.

(d) {\it Theoretical vs.~physical equivalence}: which contrasts formal equivalence with the equivalence of fully interpreted theories.

I will apply this scheme to answering questions (iii)-(iv) for gauge-gravity dualities. I will argue that the things that are physically relevant are those that stand in a bijective correspondence under duality: the {\it common core} of the two models. I therefore conclude that most of the mathematical and physical structures that we are familiar with, in these models (the dimension of spacetime, tensor fields, Lie groups), are largely, though crucially never entirely, {\it not} part of that common core. Thus, the interpretation of dualities for theories of quantum gravity compels us to rethink the roles that spacetime, and many other tools in  theoretical physics, play in theories of spacetime.

\newpage
\pagestyle{plain}

\tableofcontents

\newpage

\section*{Introduction}\label{intro}
\addcontentsline{toc}{section}{Introduction}

In their programmatic special issue on the emergence of spacetime in quantum theories of gravity, Huggett and W\"uthrich (2013:~p.~284) write that the programme of `interpreting a theory `from above', of explicating the empirical significance of a theory, is both `philosophical', in the sense that it requires the analysis of concepts, and crucial to every previous advance in fundamental physics... As such, it must be pursued by the study of theory fragments, toy models, and false theories capturing some promising ideas, asking how empirical spacetime relates to them.'

Huggett and W\"uthrich further articulate their project around three foci of attention:\\
\indent(1)~ Does quantum gravity {\it eliminate} spacetime as a fundamental structure?\\
\indent(2)~ If so, how does quantum gravity explain the {\it appearance} of spacetime?\\
\indent(3)~ What are the broader implications of quantum gravity for metaphysical accounts\\
\indent ~~~~~~of the world?\\
These are indeed central questions for candidate theories of quantum gravity. If the answer to the first question turns out to be affirmative, the impact on the philosophy of spacetime will be spectacular: for there will be, at the fundamental level, {\it no} spacetime. In this essay I consider the contribution, to the answering of questions (1) and (3), of one particular quantum gravity approach: gauge-gravity duality.\footnote{For an expository overview, see e.g.~Ammon and Erdmenger~(2015). De Haro et al.~(2016) is a conceptual review.} This is an approach developed in the context of string and M theory, but it has broader ramifications: e.g.~applications to condensed matter physics and heavy-ion collisions (Ammon and Erdmenger (2015:~III)). 
I will not consider  question (2) here.\footnote{There is a growing literature on the contribution of gauge-gravity dualities to question (2), see e.g.: 
De Haro~(2017), Dieks et al.~(2015), 
Rickles~(2012). 
}

Thus I will lay out, in the first part of the essay (Section \ref{eliminate}), the conceptual scheme for dualities that I advocate. My argument will proceed by analysing dualities (\S\ref{concdual}) in terms of four contrasts, which lead up to making the distinction between theoretical equivalence and physical equivalence. 

A duality is an isomorphism between (bare) theories. Then the four contrasts are:\\
\indent(a) {\it Bare theory vs.~interpreted theory} (\S\ref{thdual}): A {\it bare theory} is a triple of states, quantities, and dynamics, each of which are construed as structured sets, invariant under appropriate symmetries. An {\it interpreted theory} has, in addition, a pair of interpretative maps to physical quantities and other entities.
\\
\indent(b) {\it Extendable vs.~unextendable theories} (\S\ref{unext}-\S\ref{gly}): theories that do, respectively do not, admit suitable extensions in their domains of application. I will also allow for a weaker conception of `unextendable theory', according to which unextendable theories may admit an extension via e.g.~{\it couplings} to other theories in their domain, but are such that their interpretations are {\it robust}, i.e.~unchanged under such extensions.\\
\indent(c) {\it External vs.~internal interpretations} (\S\ref{interpretedth}, \S\ref{twoi}, \S\ref{4cond}): interpretations that are obtained from outside (e.g.~by coupling the theory to a second theory that has already been interpreted), respectively from inside, the theory, 
i.e.~from the role that states, quantities, and dynamics have within the theoretical structure.\\
\indent(d) {\it Theoretical vs.~physical equivalence} (\S\ref{dualtheor}): formal equivalence (i.e.~agreement of the bare theories, but with possible disagreement of the interpretations) vs.~full equivalence of the interpreted theories: i.e.~agreement of both the bare theory and the interpretive maps. 

These contrasts build upon each other: so that (a) is used in the analysis of (b); (a)-(b) are jointly used in the interpretative analysis of (c); and (a)-(c) are all needed in order to reach a verdict distinguishing theoretical vs.~physical equivalence, as (d) intends. 


My account will provide sufficient details, so that the scheme can be readily applied to other cases; and I will give several examples that will work toward applying the scheme to gauge-gravity dualities. However, 
a {\it full} account of theoretical and physical equivalence, doing full justice to the intricacies of the matter, will have to be left for the future. Further formal and conceptual development of the scheme is in De Haro and Butterfield (2018). 

Of course, not all of the above notions are completely new. But my construal of them is largely novel (the only exception being the contrast (c), for which I am in full agreement with, and just develop further, the position of Dieks et al.~(2015) and De Haro (2017)). In particular, the way I here articulate the notions of theoretical and physical equivalence in terms of the contrasts (a)-(d), so that I can successfully analyse dualities, are novel and are intended to add to the literature on both dualities and equivalence of theories.

In the second part of the essay (Section \ref{ggd}), I will apply the scheme (a)-(d) to gauge-gravity duality. This will answer the following two questions, from the perspective of this quantum gravity programme: (iii) the nature of spacetime in quantum gravity (\S\ref{sayst}), (iv) the broader philosophical and physical implications (\S\ref{metaph}). Philosophers of physics have started to address the philosophical significance of dualities in recent years: and I will compare with these works in \S\ref{compare}, so as to clarify my own contribution.

The organisation of the second part of the essay thus responds to the questions (1) and (3) posed by Huggett and W\"uthrich, taken as specifically about gauge-gravity duality. 

Gauge-gravity duality is one particular approach to quantum gravity. Briefly, it is the equivalence between: 

(I: Gravity) On the one hand: a theory of quantum gravity in a {\it volume} bounded inside a certain surface.

(II: QFT)  On the other: a quantum field theory defined on that {\it surface}, which is usually, in most models, at ``spatial infinity'', relative to the volume. 

I will argue that, despite the apparent innocence of the references to spacetime appearing in this brief summary of the duality: the physical interpretation of this duality calls for a revision of the role that most of our physical and mathematical concepts are supposed to play in a fundamental theory---most notably, the role of spacetime. And the interpretation of the duality itself requires us to carefully reconsider the philosophical concepts of theoretical equivalence and physical equivalence.

\section{Theories, duality, and physical equivalence}\label{eliminate}

In this Section, I briefly describe the scheme (a)-(d), which I will apply in Section \ref{ggd}.
In \S\ref{thdual}, I specify more exactly what I mean by `theory' and related notions. In \S\ref{concdual}, I give my conception of a duality between such theories. \S\ref{dualtheor} describes how, for theories ``of the whole world'', duality is tantamount to physical equivalence, i.e.~the theories at issue being really the same theory: this will involve two conditions (\S\ref{4cond}). In \S\ref{gly}, I compare my account with Glymour's notion of equivalence.

\subsection{The conception of a theory}\label{thdual}

Before we engage with the interpretation of dualities (which we will do in \S\ref{concdual}-\S\ref{dualtheor}), we need to have conceptions of theory that are sufficiently articulated that they make an analysis of physical equivalence possible. I first introduce, in \S\ref{intth}, the notion of a {\it bare theory}. Then, in \S\ref{interpretedth}, I discuss the notion of an {\it interpreted theory}. 

\subsubsection{Bare theory}\label{intth}

I take a {\it bare theory} to be a triple $T=\bra{\cal H},{\cal Q}, \cal D\ket$ comprising: (i) a set ${\cal H}$ of states, endowed with appropriate structure; (ii) a set of physical quantities ${\cal Q}$, endowed with appropriate structure; (iii) a dynamics ${\cal D}$, consistent with the relevant structure. Such a triple will generally also be endowed with {\it symmetries}, which are automorphisms $s:{\cal H}\rightarrow{\cal H}$ preserving (a subset of) the valuations of the physical quantities on the states (for details, see De Haro and Butterfield~(2018:~\S3.3)), and which commute with (are suitably equivariant for) the dynamics $\cal D$. 

For a {\it quantum} theory, which will be our main (though not our sole!) focus: we will take ${\cal H}$ to be a Hilbert space; ${\cal Q}$ will be a specific subset of operators on the Hilbert space; and $\cal D$ will be taken to be a choice of a unique (perhaps up to addition by a constant) Hamiltonian operator from the set ${\cal Q}$ of physical quantities. In a quantum theory, the appropriate structures are matrix elements of operators evaluated on states; and the symmetries are represented by unitary operators. But as mentioned: the present notion of a bare theory applies equally well to classical and to quantum theories, and there will be {\it no} requirement that a duality must relate quantum theories.

A theory may contain many more quantities, but it is only after we have singled out the ones that have a physical significance that we have a {\it physical}, rather than  a {\it mathematical}, theory or model. The quantities ${\cal Q}$, the states ${\cal H}$, and the dynamics $\cal D$ have a physical significance at a possible world $W$, and within it a domain of application $D_W$ (which we can think of as a subset $D_W\subseteq W$), though it has not yet been specified what this significance may be, nor what the possible world ``looks like''.\footnote{This framework allows of course for models that do not describe actual physical reality. While one can give various metaphysical construals to concepts such as `possible worlds', I here need only a minimal conception of possible worlds as ``how things can be'', in the context of a `putatively fundamental' spacetime physics. Thus I do not mean to subscribe to a specific metaphysics of possible worlds.} To determine the physical significance of the triple, a physical interpretation needs to be provided: which I do in the next subsection. 

So, I will dub as the {\it bare theory}: just the formal triple $T=\bra{\cal H},{\cal Q},\cal D\ket$, together with its structure, symmetries, and rules for inferring propositions, such as: `the value of the operator $Q\in {\cal Q}$ in the state $s\in{\cal H}$ is such and such'. But there is no talk of empirical adequacy yet.

We normally study a theory through its models. A {\it model} is construed as a representation of the theory, the triple $\bra{\cal H},{\cal Q},\cal D\ket$,\footnote{`Representation' is here meant in the mathematical sense of a homomorphism. In this paper, I will only consider isomorphic representations: for the non-isomorphic case, see De Haro and Butterfield (2018:~\S2.2.2-\S2.2.3). This reference argues that a more general account of a model can be had: as an instantiation, or a realization, of a theory.} and will be denoted by $M$. A model $M$ of a theory $T$ may include, in addition to the triple, some variables that are part of the descriptive apparatus but have no physical significance (in the sense of the last but one paragraph) from the point of view of the theory: I will call this the {\it specific structure} of the model $M$. 
Since a theory can be represented by any of its models, my account does not seek to eliminate the specific structure, but to identify the core $\bra{\cal H},{\cal Q},\cal D\ket$ of the models: as that structure that is preserved across equivalent models (see \S\ref{concdual}). 
In our duality of interest, gauge-gravity duality, I will treat the two sides of the duality as two models of the theory.

Notice that the notion of model I use here, as a representation (more generally, an instantiation) of a theory, is distinct from the more common characterisation of a model as a specific solution for the physical system concerned (e.g.~a specific trajectory in the state space, or a possible history, according to the equations of the theory): which I indeed here reject. My notion is motivated by dualities because they prompt us to move what we mean by a theory ``one level up'', in abstraction: and I accordingly take what I mean by a model ``one level up'', keeping the relation between theory and model fixed, rather than introducing a new level between theory and model. Thus, what used to be `two distinct theories', each of them having their own models, are now `two models of a single theory', each model having its own set of solutions, related to each other by the duality. What dual models have in common, i.e.~their {\it common core}, is isomorphic between the models. For more on the motivation for, and the uses of, these notions of theory and of model, see De Haro and Butterfield (2018:~\S2). 

\subsubsection{Interpreted theory}\label{interpretedth}

There is a certain minimalism to the presented conception of theory: since in scientific practice one must be able to tell, in a given experiment or physical situation to which the theory is supposed to apply, what the relevant quantities are that correspond to the empirical data. The above specification of a theory as a triple makes no reference as yet to this: only the existence of some such relation, for a class of possible worlds $W$, is assumed. So, when interpreting a theory, one wishes to also:

(0) establish the meaning of certain theoretical entities (if one is a realist), whether directly measurable or not;

(1) establish some kind of bridge principles between the physically significant parts of the theory and the world. 

These two desiderata will be fulfilled by the two interpretative maps (denoted $I^0,I^1$ below). Furthermore, one may also wish to establish {\it theoretical principles} that, for example, interconnect various experimental results (symmetry and locality being just two examples of such theoretical principles often considered in physics).

So, I take a {\it physical interpretation} to be a pair of partial maps, preserving appropriate structure, from the bare theory to some suitable set of physical quantities. I will denote the maps as $I_T:=(I^0_T,I^1_T):T\rightarrow D_W$, where $T$ is the triple\footnote{Or Cartesian products thereof: e.g.~in quantum mechanics, the interpretation map maps e.g.~expectation values to real numbers in the world. The expectation values themselves are maps from Cartesian products of states and quantities to real numbers.} and $D_W$ is the domain within the possible world $W$ (a pair of possible worlds, since there are two maps: more on this below; for simplicity in the notation, I will often drop the subscript). 
The restriction to only a domain within a possible world stems from the wish to account for theories that do not describe the whole world, but only part of it. The requirement that the map needs only be partial means that not all the elements of the theory necessarily refer. The more adequate the theory, the more elements will refer and the map will satisfy additional conditions, but we will not use this here.

The first partial map, $I_T^0$, is from the triple, $\cal H,{\cal Q}$ and $\cal D$, to the quantities understood conceptually (potential energy, magnetic flux, etc.), i.e.~it corresponds to the theoretical meaning of the terms. Of course, this theoretical meaning is informed by experimental procedures, and it will usually involve the quantities as realized in all possible laboratory experiments, through observations, or even through relations between experiments and-or observations, in a domain $D^0$ of a possible world $W^0$. But this possible world is not to be understood as our world in its full details of context, but as a conceptual realm that fixes the reference of the theoretical concepts.\footnote{Cf.~De Haro and Butterfield (2018:~\S2), where the first map is an intension, and the second an extension.}

The second partial map, $I^1_T$, is from the triple, $\cal H,\cal Q$ and $\cal D$ (or Cartesian products thereof), to the set of values (the set of numbers formed by all experimental or observational outcomes, as well as the relations between them), i.e.~the domain $D^1$ to which the theory applies, within a possible world $W^1$. That set of values is typically (minimally) structured, and such minimal structure is to be preserved by the map. Thus typically, the second map maps e.g.~individual states or quantities, e.g.~by pairing them: $I^1_T:\cal H\times\cal Q\rightarrow\mathbb R$, where $\mathbb R$ is endowed with addition and multiplication (for a concrete example, see \S\ref{twoi}), and likewise for the dynamics. Notice that, unlike the first map, the second map depends on contingent facts about the possible world $W^1$, such as the context of a measurement.

The codomains of $I^0_T$ and $I^1_T$ as thus defined might, at first sight, seem to be again just theoretical or mathematical
 entities, themselves in need of interpretation---a space of functions in the first example, a set of real numbers endowed with addition and multiplication, in the second. But this is not what is intended. We are to think of the codomain of $I^1_T$ as representing the real world in a straightforward way: so, for example, the reals measuring the energy represent the value of a quantity in appropriate units, given by the position of a voltimeter's pointer on a scale (see some examples in \S\ref{twoi}). The point is to map from theories to structured sets of functions and numbers, which do {\it not} describe more theory, but rather are identified with a set of possible physical situations, experimental outcomes, or observational and conceptual procedures. 

In what follows, the distinction between $I^0$ and $I^1$  (and between the idealized world $W^0$ and the concrete world $W^1$, to which they map) will not be important, and so I will gloss over the `0 vs.~1' contrast: and simply talk of a pair of interpretation maps $I$ from the theory to the world, $W$. The distinction between the two cases is brought to bear in De Haro and Butterfield (2018:~\S2-3).

In the above discussion, I have so far assumed that the interpretative maps are from the elements of {\it the triples themselves}, rather than from their models, to quantities etc.~in the world. We will call such an interpretation {\it internal}: it requires nothing but the theory. But, as I will discuss in detail in \S\ref{twoi}, there are often good reasons to pursue an interpretation that is obtained by e.g.~coupling the theory $T$ to some other theory $T_{\tn{meas}}$ that is already interpreted. $T$ thus inherits its interpretation from $T_{\tn{meas}}$; and the coupling of $T$ to $T_{\tn{meas}}$ will often differ for the different models $M$ of $T$. In other words, $T_{\tn{meas}}$ may be coupled to $M$ through $M$'s specific structure. So, we  distinguish:\\
\indent(1)~~{\it External interpretation of $T$}: a pair of maps, as above, $I_M:M\times T_{\tn{meas}}\rightarrow D_W$, from the {\it model} $M$ of the theory $T$, coupled to the theory of measurement $T_{\tn{meas}}$ (or some other relevant theory that provides an interpretation), to the domain $D_W\subseteq W$.\\
\indent(2)~~{\it Internal interpretation of $T$}: a pair of maps, as above, $I_T:T\rightarrow D_W$, from the {\it theory}, $T$, to the domain, $D_W\subseteq W$. 

I will call a bare theory, once it is equipped with an interpretation, the {\it interpreted theory}.
It is the physical interpretation that enables the theory to be empirically successful and physically significant. 

The interpretation maps involve, of course, philosophically laden issues. But the aim here is not to settle these issues, but rather to have a scheme in which the formal, the empirical, and the conceptual are clearly identified and---as much as possible---distinguished. The interpretational scheme should make the ontological commitments {\it explicit} (more on this in \S\ref{gly}). 

The fact that the first map, $I^0_T$, is conceptual, and that the two maps also respect relations between quantities (which might not be directly and independently measurable, while still being physically significant), reflect the fact that the set of quantities ${\cal Q}$ need not be restricted to measurable or observable quantities. I will illlustrate this with an example in \S\ref{gly}. 

The above formulation of a bare theory as a triple is minimalist. But, with the interpretation maps added, it is strong enough---because of the complete specification of the set of physical quantities---that, under two conditions mentioned below, it will determine whether two theories are about the same subject matter.
Questions concerning the identity of two such triples will be questions concerning the sameness of models (\S\ref{4cond}), rather than standard cases of underdetermination of theory by empirical data. This is because we asssume that the triple $T=\bra {\cal H}, {\cal Q}, {\cal D}\ket$, together with the valuations constructed from the syntax, is well-defined and consistent, and that it encompasses all the empirical data (and relations between the data), in a certain domain $D_W$ within $W$. The latter condition can be restated as the requirement that the interpretative map be {\it surjective}, so that no element of the domain $D_W$ is left out from the model's description.\footnote{The qualification, `in a domain within a possible world', is important because the theory need not be complete at {\it our world}. A model may, for example, be complete within a given range of parameters not containing the relevant values for our world. This is why completeness is to be construed as relative to $D_W$ and $W$.\label{PW}} I will call such a model {\it complete}.  Thus, the equivalence of two triples means that the theories agree about everything they deem physical. On the other hand, in standard cases of underdetermination considered in the philosophy of science, the theories are underdetermined by what is measurable or observable, by the available evidence, or even by the actual empirical data obtained by scientists.
The completeness of the theory is thus a necessary condition for duality to which I return.

\subsection{The conception of duality}\label{concdual}

With the conception of a theory considered in \S\ref{thdual}, a {\it duality} is now construed as an {\it equivalence of theories}. More precisely, it is an isomorphism $d:M_1\rightarrow M_2$ between two models $M_1$ and $M_2$ of a theory $T$: there exist bijections between the models' respective sets of states and of quantities, such that the values of the quantities on the states are preserved under the bijections. (In the case of quantum theories, these values are the set of numbers $\bra s_1|Q|s_2\ket$, where $s_1,s_2\in{\cal H}$, $Q\in{\cal Q}$.) The duality is also required to commute with ($d$ is to be equivariant for) the two models' dynamics, and to preserve the symmetries of the theory.


The notion of duality in this subsection is motivated by both physics and mathematics. Duality in mathematics is a formal phenomenon: it does not deal with physically interpreted structures (even though, of course, several of the mathematical dualities turn out to have a physical significance). But this is also how the term is used by physicists: it is attached to the equivalence of the formal structures of the theories, regardless of their interpretations.

Duality, as a formal equivalence between two triples without the requirement of identical interpretations, is thus a special case of theoretical equivalence: an isomorphism, as above.

Like the conception of a theory, my conception of a duality is minimalist. On this definition, for instance, position-momentum duality in quantum mechanics is indeed a duality. The duality has two models, namely the formulations of quantum mechanics based on, respectively, the $x$- and the $p$-representations of the Hilbert space: Fourier transformation being the duality map. This duality is, of course, somewhat trivial, because the two models contain {\it the same amounts of specific structure}, in the sense above: in short, a single variable. And indeed, I regard it as a virtue that my conception of duality is general enough that both familiar, and relatively simple, dualities, as well as the more sophisticated ones in quantum field theory and quantum gravity, all qualify as dualities, under the same general conception. Indeed, I take it that: 

{\it One of the lessons of duality is that ``widely differing theories'' are (surprisingly) equivalent to each other, in the same sense of equivalence in which two `notational variants' differ from each other.} 

An important question, given the conceptions of theory, model, and duality I have introduced, is whether one can verify, in some well-understood examples, that the scheme leads to the correct verdicts regarding duality. Examples of equivalences between very different-looking quantum field theories are provided in De Haro and Butterfield (2018), where it is shown how duality thus construed obtains. The focus of this essay---in line with the quote by Huggett and W\"uthrich given in the Introduction---is in applying this conception, and the four contrasts (a) to (d), to shed light on {\it duality in quantum gravity.} 

The two-pronged conception of an interpreted theory as a triple plus an interpretation, together with the notion of duality as isomorphism, allow us to introduce the notion of the {\it physical equivalence} of models. The discussion of duality so far indeed prompts us to distinguish {\it theoretical equivalence} from {\it physical equivalence}: the latter being the complete equivalence of two models as descriptions of {\it physical systems}, i.e.~models with identical interpretations. The difference may be cashed out as follows: theoretically equivalent models, once interpreted, ``say the same thing'' about possibly {\it different subject matters} (different parts of the world), whereas physically equivalent models say the same thing about the {\it same subject matter} (the same part of the world).  
There will also be a weak, but interesting, form of physical equivalence, in which two dual models describe a single given world equally well, even if in other cases they may also describe different worlds. More on this in \S\ref{dualtheor}.

Duality, then, is one of the ways in which two models can be theoretically equivalent, without its automatically implying their physical equivalence. For instance, a duality can relate a real and an imagined or an auxiliary system. In such a case, duality is a useful and powerful calculational device---and nothing more. But it is, of course, those cases in which dualities do reveal something about the nature of physical reality, that prompts the philosophical interest in dualities: cases in which the interpretation of the duality promotes it to physical equivalence.

\subsection{From theoretical equivalence to physical equivalence}\label{dualtheor}

Having introduced, in the previous two subsections, my conception of duality, and the four contrasts (a) to (d) mentioned in the Introduction, we now come to the central question in this Section: {\it When does duality amount to physical equivalence?} I first discuss, in \S\ref{twoi}, the external and internal interpretations of a theory (already briefly introduced in \S\ref{interpretedth}). In \S\ref{4cond}, I state two conditions for physical equivalence: internal interpretation and unextendability. In \S\ref{unext}, I discuss the physical equivalence of dual theories that are unextendable. 

\subsubsection{External and internal interpretations of a theory}\label{twoi}

In this subsection, I further develop the external and internal interpretations, and in particular two cases: (i) cases of external interpretations, in which {\it physical equivalence fails to obtain}, despite the presence of a duality; (ii) cases in which an external interpretation is not consistently available (where `consistently' will be qualified below), so that one can only have an internal interpretation, and hence there is {\it physical equivalence}.

Let me illustrate the external interpretation with an elementary example that should make clear the difference between duality as a case of theoretical equivalence, and physical equivalence. 
Consider classical, one-dimensional harmonic oscillator ``duality'': an automorphism, $d:{\cal H}\rightarrow\cal H$, defined by $d:{\cal H}\ni(x,p)\mapsto({p\over m\o},-m\o x)$, from one harmonic oscillator state to another, leaving the dynamics $\cal D$ invariant---namely, the Hamiltonian $H={p^2\over2m}+\half kx^2$ and the equations of motion that $H$ defines. So, it is an automorphism of $T_{\tn{HO}}=\bra {\cal H},{\cal Q},\cal D\ket$.\footnote{I have described the states and dynamics of the harmonic oscillator: the quantities ${\cal Q}$ include e.g.~any powers (and combinations of powers) of $x$ and $p$.}
But this automorphism of $\bra {\cal H},{\cal Q},\cal D\ket$ does not imply a physical equivalence of the states:\footnote{In a world consisting of a single harmonic oscillator {\it and nothing else}, the two situations could not be distinguished, and one might invoke Leibniz's principle to identify them. This amounts to adopting an internal interpretation, in the sense of \S\ref{interpretedth}.} the two states are clearly distinct and describe different physical situations: since the map $d$ relates an oscillator in a certain state of position and momentum, to an oscillator in a {\it different} state. 

This difference is shown in the fact that there is an independent way to measure the `position' of the oscillator at a given time: one sets the oscillator and a standard rod side by side, observes where on the rod the oscillator is located, and so carries out a measurement of the former's position. We can picture this as coupling harmonic oscillator theory, $T_{\tn{HO}}$, to our theory of measurement $T_{\tn{meas}}$, and interpret the measurement as measurement of the oscillator position. I will call such an interpretation of $T_{\tn{HO}}$ an {\it external interpretation} (cf.~\S\ref{interpretedth}). It is obtained by inducing the interpretation of $T_{\tn{HO}}$ from an already interpreted theory $T_{\tn{meas}}$, or by extension to $T_{\tn{HO+meas}}$. And I call a theory, that can be coupled or extended in this way, an {\it extendable} theory. 

But there are cases---such as cosmological models of the universe, and models of unification of the four forces of nature---in which these grounds for resisting the inference from duality to physical equivalence---a resistance based on the possibility of finding an external theory $T_{\tn{meas}}$---are {\it lost}. For the quantum gravity theories under examination---even if they are not {\it final} theories of the world (whatever that might mean!)---are presented as candidate descriptions of an {\it entire} (possible) physical world: let us call such a theory $T$.\footnote{These are candidate descriptions of {\it possible} worlds, rather than the actual world. For example, the models which we will consider in Section \ref{ggd} entail a negative cosmological constant, whereas our universe seems best described by a {\it positive} cosmological constant. But the interest in such models is, of course, that: (1) given their rarity, {\it any} consistent four-dimensional theory of quantum gravity is interesting; and (2) such idealised models contain helpful lessons for the case of a positive cosmological constant.\label{cosmol}
} So, there is {\it no} independent theory of measurement $T_{\tn{meas}}$ to which $T$ should, or could, be coupled, because $T$ itself should be a closed theory (an {\it unextendable} theory: see \S\ref{unext}). In the next subsection, I will use these ideas to spell out the conditions under which two dual models are physically equivalent.


\subsubsection{Internal interpretation and unextendability allow sameness of reference, and so physical equivalence}\label{4cond}

We return to dualities as isomorphisms of models (cf.~\S\ref{concdual}); and so, we consider, specifically, the interpretation of two models, $M_1$ and $M_2$, ``of the whole world''. In this Section, I propose a condition that will, together with the internal interpretation, secure physical equivalence, in the sense that one is justified in taking duals to be physically equivalent: and I give the arguments to that effect. 

The leading idea of an internal interpretation is that the interpretation has not been fixed a priori, but will be developed starting from the duality. (Or, if by some historical accident, an interpretation has already been fixed, one should now be prepared to drop large parts of it.) 
The requirement that, I propose, justifies the use of the internal interpretation such that uniqueness of reference is secured, is as follows:\\
\\
{\it Unextendability}: roughly, `the interpretation cannot be changed by coupling the theory to something else or by extending its domain'. Unextendability replaces the somewhat vague phrase `of the whole world' in the previous paragraph, and I will expound it in \S\ref{unext}. \\

Unextendability plays a key role in inferring physical equivalence. For it ensures that there is ``no more to be described'' in the physical world, and that the models cannot be distinguished, even if their domains of application were to be extended (since no such extension exists). And so, it ensures that the internal interpretation {\it can be trusted} as a criterion of physical equivalence (cf.~\S\ref{unext}), as I now argue. 

Starting, then, from two such dual models, $M_1$ and $M_2$ of $T$, the duality map lays bare the invariant content $\bra{\cal H},\cal Q, D\ket$, as that content which is common to $M_1$ and $M_2$, through the duality map (cf.~\S\ref{interpretedth}). This is the starting point of the {\it internal interpretation}, for both the theory and the models. I now propose that an internal interpretation of a theory, satisfying the two stated conditions, is the same for the two models (in the sense of \S\ref{concdual}), and in particular its reference is the same:

(i) the formalisms of the two models say the same thing: for they contain the same states, physical quantities and dynamics (i.e.~the domain of the maps is identified by the isomorphism), and (ii) their physical content is also the same: for the interpretation given to the physical quantities and states is developed from the duality {\it and nothing else}: so, the codomains of the maps are the same, and they coincide with the entire world. I am thus here proposing that {\it the domains of the worlds described by two dual models, and the worlds themselves are the same}: 
this is because, on an {\it internal} interpretation, the two worlds, in all their physical facts, are `constructed, or obtained from', the triples. 

There is a way in which this inference, from dual models with internal interpretations, to identical worlds, might fail: there might be more than one internal interpretation, and therefore more than one codomain $D_W$ described by the theory. For in that case, despite the isomorphism of the two models, one might be tempted to think that one model could be better interpreted in one way, and the other better interpreted in another way.


But I take this objection to be misguided: the point is that, even if there is more than one internal interpretation, the reference of a given internal interpretation is the same for any two isomorphic models. Remember that, by definition, an internal interpretation cannot discern between models, because it starts from the theory, as a triple, {\it and nothing else}. So, there can be no reason for the interpretation to distinguish one model from the other---they both describe the world equally well and in the same way, according to that internal interpretation. In other words: {\it even if a single common core admitted several internal interpretations, each of them would refer to a single possible world, which would be the single reference of the corresponding internal interpretation of all the models isomorphic to the common core.}

Let me spell out in more detail this inference from the isomorphism of models, to the identity of the internal interpretations and identity of the worlds described (hence physical equivalence), under the unextendability condition. There are two ways to make this inference. The first argument, from the unextendability of the models, will be given in \S\ref{unext}: it is a version of Leibniz's principle of the identity of indiscernibles: which applies here, because the two models describe the entire world. The second argument, given in the previous paragraph, simply follows from the definition of the internal interpretation, given in \S\ref{interpretedth}: an internal interpretation is constructed from the triple of the theory {\it and nothing else} (so, the specific structure of a model is not to find a counterpart in the world, since the interpretation must be invariant under the duality map). Thus, given an internal interpretation of the theory, the codomain of that interpretation, mapped from the two models, is the same by definition: {\it since the internal interpretation is insensitive to the differences in specific structure between the models, its reference must be the same.}\footnote{The internal interpretation $I$ is a partial surjective map from the theory to the world. But using the forgetful map from the model to the theory (the map which strips the model of its specific structure), we can construct an internal interpretation of the model, as the pullback of the interpretation map $I$ by the forgetful map. It is in this sense that I here speak of internal interpretations of the models as well as of the theory.} 
Explicitly, $I_{M_1}=I_{M_2}\circ d$, where $d:M_1\rightarrow M_2$ is the duality map. Thus, such thorough-going dualities can be taken to give {\it physical equivalence} between apparently very different models.

It is important to note that this second argument for the identity of the codomains follows from the {\it definition} of an internal interpretation, given in \S\ref{interpretedth} (together with the two stated conditions). What is surprising about a duality that is a physical equivalence, then, is not so much that two very different models describe {\it the same world}, but rather that {\it there is an internal interpretation} to be constructed from such minimal data as a triple:\footnote{I call the triple `minimal' data because it does not contain specific structure, which we normally think of as giving a model, and its interpretation, its particular features. So, the internal interpretation constructed from a triple may be rather abstract: yet, the claim is that it is an entire world!} and so, what is surprising is that there is a (rich) world for such a triple to describe! Admittedly: it would be hardly surprising if the internal interpretation described something as simple as the real line. But, in the examples we are concerned with here, the internal interpretation describes far richer worlds!

I do not claim to have established physical equivalence, under the conditions of internal interpretation and unextendability, as a matter of logical necessity, just from the notions of theory, model, and interpretation. Doing so would require a deeper analysis of the notion of reference itself, and the conditions under which it applies to scientific theories: which is beyond the scope of this paper (cf.~e.g.~Lewis (1984)). What I claim to have argued is that, making some natural assumptions in particular about how terms refer in ordinary language and in scientific theories, the use of the internal interpretation and unextendability does deliver physical equivalence.


\subsubsection{Unextendability, in more detail}\label{unext}

I turn to consider theories, such as theories of quantum gravity, $T_{\tn{QG}}$ say, for which there is no extra physics to which $T_{\tn{QG}}$ can be coupled or extended. Being a description of the entire physical universe, or of an entire domain of physics, I will take the interpretation $I_{\tn{QG}}$ to be {\it internal} to $T_{\tn{QG}}$. Thus, as a sufficient condition for being justified in the use of an internal interpretation, I have required, in \S\ref{4cond}, that $T_{\tn{QG}}$ be an unextendable theory. An interpretation map $I_{\tn{QG}}$ only requires the triple $T_{\tn{QG}}=\bra {\cal H},{\cal Q},D\ket$ as input, and it only involves the triple's elements and their relations---it does not involve coupling $T_{\tn{QG}}$ to other theories. In such a case, duality preserves not only the formalism, but necessarily also the structure of the concepts of two complete and mathematically well-defined models: if one model is entirely self-consistent and describes all the relevant aspects of the world, then so must the other model. And so, duality becomes physical equivalence. Thus, in other words, we are really talking about different formulations of a {\it single theory}. 

Let me spell out the (sufficient) condition, suggested by this discussion, for a theory to admit an internal interpretation, since it will be important in \S\ref{gly}. A bare theory $T$ in a domain $D_W$ of a possible world $W$ is {\it unextendable} iff:

(i)~~$T$ is a complete theory in the domain of applicability $D_W$ at $W$;

(ii)~There is no other theory $T''$ for the possible world $W$ (or another possible world that includes it) and domain $D_W$, such that: for some $T'$ isomorphic to $T$,  $T'\subset T''$ (proper inclusion).\footnote{$T'$ is a fiducial theory that may well be {\it identical} to $T$. But in general, it may be the case that $T\subset T''$ is not true but $T\cong T'\subset T''$ is. In other words, $T\subset T''$ may only be true up to isomorphism.}

(iii)~The domain of applicability $D_W$ coincides with the world described, i.e.~$D_W=W$.

Note that the possible world $W$ is fixed by the interpretation. Unextendability is thus a relation between bare theories and worlds, and is thus a property of {\it interpreted theories}.\\

Recall the notion  of completeness of a theory, in (i), introduced in \S\ref{intth}: as well-defined, consistent, and encompassing all the empirical data in a certain domain $D_W$ (i.e.~a partial surjective map). Condition (ii), in addition, requires that there is no extension of the theory at $W$: or, in other words, the theory already describes all the physical aspects of the relevant domain at $W$. 
Since the relation of isomorphism in (ii) is formal, (ii) is a sort of ``meshing'' condition between (i)---or, more generally, between the idea of ``not being extendable''---and the formal relation of isomorphism between bare theories.\footnote{I have argued that unextendability is a sufficient, though not a necessary, condition for the coherence of an internal interpretation. The condition is not necessary because one can envisage a theory (e.g.~general relativity without matter) receiving an internal interpretation (e.g.~points are identified under an active diffeomorphism, taken as the lesson of the hole argument). This interpretation does not change when we couple the theory to matter fields: and I will say that such an internal interpretation is {\it robust} against extensions. If all possible extensions of a theory preserve an internal interpretation, then such an interpretation is justified. If the extensions suggest diverging interpretations, then one needs to specify the domain of the extension before one is justified in interpreting the theory internally. In other words, (i)-(iii) can be weakened, if what we want is not sameness of reference but of descriptive abilities.\label{unextendability}} Condition (iii) is the usual condition that $T$ is a theory ``of the whole world'', i.e.~its domain of application is the entire world. Notice that (i) and (iii) do not suffice for physical equivalence: one needs something like the technical requirement (ii) (see also the next Section).

I have concluded that, on an internal interpretation, there is no distinction of content between two dual models. In \S\ref{gly}, I will address two different {\it purposes} for which the distinction between two dual models is irrelevant, besides the ontological purposes so far discussed: viz.~the logical and empirical purposes. This is not to deny that there are other significant---metaphysical, epistemic, and pragmatic---purposes or uses of physical theories, for which the differences are significant. For instance, one of the main pragmatic virtues of gauge-gravity dualities is that one theory is tractable in a regime of values of the parameters where the other theory is intractable.

\subsection{Comparison with Glymour's notion of equivalence}\label{gly}

How does the conclusion, that the two theories related by gauge-gravity duality admit internal interpretations, and that under two additional conditions, duality implies physical equivalence, compare with the relevant philosophical literature on equivalence of physical theories? To discuss this, I will recall the usual strategy by which, faced with apparently equivalent theories, physicists try to break the equivalence; and relate this to an influential discussion, by Glymour. I will agree with Glymour's verdicts for his examples, but I will argue that this depends on the theories in the examples being {\it extendable}.

It is a commonplace of the philosophy of science that, confronted with theoretically inequivalent, but empirically equivalent theories, physicists naturally imagine resorting to some adjacent piece of physics that will enable them to confirm or disconfirm one of the two theories as against the other. The classic case is: confronted with differing identifications of a state of rest in Newtonian mechanics, Maxwell proposes a measurement of the speed of light. There is a parallel for dualities: when two theories are both theoretically and empirically equivalent, we can still argue, by an extension or by a resort to some adjacent piece of physics, for their physical inequivalence. This is articulated in the contrast, in \S\ref{unext}, between extendable and unextendable theories.

Glymour's (1977) discussion of equivalence of theories uses the syntactic conception of a theory as a set of sentences closed under deducibility. He introduces the notion of `synonymy': two theories are synonymous when they are, roughly speaking, logically equivalent. That is, there is a well-defined inter-translation between them.\footnote{Technically, what is required is a common definitional extension. 
Barrett and Halvorson (2016:~\S4,~Theorems~1-2) show that Glymour's `synonymy', i.e.~there being a common definitional extension, is equivalent to an amendment of Quine's `translatability'. }
Although my use of theories as triples puts me closer to the semantic conception (the syntactic conception's traditional rival), in fact Glymour's criterion of synonymy meshes well with my notion of a duality, construed as an isomorphism of triples, equipped with rules for forming propositions about e.g.~the value of a quantity or a state. One considers the set of well-formed sentences built from two triples $T_1$ and $T_2$, e.g.~statements of the type `the value of the quantity $Q_1\in{\cal Q}$ (resp.~$Q_2$), in such and such state, is such and such'. Duality then amounts to isomorphism between two such sets of sentences. And this is a case of synonymy in Glymour's sense.

But does this immediately lead to physical equivalence? No. And the reasons provided in \S\ref{unext} are similar to the ones Glymour gives. He envisages theories that  are synonymous, in the sense just described: yet are not physically equivalent.
Recall Glymour's thought experiment (p.~237):

\begin{quote}\small
Hans one day announces that he has an alternative theory which is absolutely as good as Newtonian theory, and there is no reason to prefer Newton's theory to his. According to his theory, there are two distinct quantities, gorce and morce; the sum of gorce and morce acts exactly as Newtonian force does.
\end{quote}

Glymour denies that the Newtonian `force theory' and Hans' `gorce-and-morce theory' are physically equivalent. He argues that they are empirically equally adequate, but not equally well {\it tested}. His reasons for this are, partly, ontological (``I am, I admit, in the grip of a philosophical theory'', p.~237), and his ontology leads him to prefer the Newtonian theory: the gorce-and-morce theory contains two quantities rather than one, but there is no evidence for the existence of that additional quantity. The argument is from parsimony: he prefers a sparse ontology. And so, Glymour's reply to Hans seems to entail that the two theories are not even theoretically equivalent, because the gorce-and-morce theory has one more quantity than the force theory.  

But even if the two theories {\it were}, or could be made, theoretically equivalent, I would still agree with Glymour's verdict of a lack of physical equivalence, in so far as one is concerned with theories that admit external interpretations, and this for two reasons: 

(a) His examples deal with classical theories of gravitation, i.e.~{\it extendable} theories admitting external interpretations, which can indeed vary widely. That these theories are extendable, beyond a certain regime of energies, can be seen from both Newton's and Einstein's theories' predictions of {\it singularities}, which are usually taken to be unphysical. 

(b) The force theory and the gorce-and-morce theory are empirically equivalent on a restricted domain, but their extensions are {\it not}: 

\begin{quote}\small
To test these hypotheses, the theory must be expanded still further, and in such a way as to make the universal force term [read instead: `gorce'] determinable (ibid, p.~248). 
\end{quote}

But, as I argued before: under the conditions stated in \S\ref{4cond}, in particular in cases in which the theory already contains all the physics it can and should contain---in case the theory is 
unextendable---such extensions are simply not given and the inequivalence does not follow. In such a case, no further relevant theory construction could tell force apart from gorce and morce. The latter phrase then surely does not refer to anything independent  and distinct from what is meant by `force', and the two theories {\it are} physically equivalent. In other words: on an internal interpretation of a theory, Glymour synonymy leads to physical equivalence.

This also illustrates a point that I mentioned in the last but one paragraph of \S\ref{interpretedth}: namely, that the set of quantities ${\cal Q}$ need not be restricted to measurable quantities. Indeed, for an unextendable theory, we can perfectly well add gorce and morce as distinct quantities to the set of quantities ${\cal Q}$, even if no measurement could possibly distinguish them. In other words, the difference of gorce and morce can be added as a quantity, whose value no empirical data can determine, in an unextendable theory.

Of course, Glymour's argument for the distinctness of the two theories has an ontological component: while Newton postulates one quantity, Hans postulates two. We assumed that we already knew which terms in each sentence referred to {\it some} things in the world (perhaps without yet knowing {\it which} things), on an external interpretation. Hans' theory was interpreted as saying that {\it two} things exist instead of just one, and this implied the inequivalence of the two theories---at least, if we assume that the interpretation maps refer as indicated in \S\ref{4cond}. To explain how this is possible, given that the theories were Glymour-synonymous, one envisaged extending the theory, thus giving an independent account of what these terms refer to: an account of what the existence of these two things would imply, upon formulation of the theory on a larger range of validity within its domain. 

That the issue at stake here is not only the theoretical equivalence of the triples, but also their physical equivalence, can be seen as follows. Glymour's reply to Hans is that he is committed to two quantities, gorce and morce, rather than a single force: and so, one might be tempted to say that, on my account, this implies that the two theories contain different quantities ${\cal Q}$: they are different, as triples, and so not theoretically equivalent. But I don't think this is necessary: in Hans' defence, his claim could be taken to be that gorce and morce are two distinct quantities related by a symmetry: and that this symmetry should lead us to identify his theory with a theory of a single quantity. On this charitable account, Hans is saying that a theory with two quantities and a symmetry relating them is the same as a theory with a single quantity, i.e.~he construes symmetries as equivalence relations, and instructs us to mod out theories by such symmetries. So, he {\it is} construing the two theories as theoretically equivalent. But Hans' improved argument again fails to produce physical equivalence, even if he claims that there is theoretical equivalence: the theories are extendable, and so, an extension of the theory beyond its domain may reveal that the purported symmetry is broken and the triples are distinct (see two paragraphs below). 

But on an internal interpretation, we cannot assume we possess an account of what `force' and `gorce and morce' mean, from outside the theory. The impossibility of an extension, therefore the lack of an independent account of what those terms mean, implies that we should not make such ontological claims {\it independently of} the equivalence of the two theories. Because the two theories are Glymour-synonymous, and there is no extension, they can be taken to be physically equivalent: and so, there are not two quantities but just one.\footnote{In discussing the internal interpretation, I am assuming that one's formulation of the theory is sufficiently perspicuous (e.g.~as a triple), that the physical quantities can be read off from it. If this is not so, it might be equally natural to say that there is no fact of the matter about whether one is committed to one or two quantities---or that such facts are underdetermined by the relevant physical quantities. Cf.~the discussion, just below, of effective field theory.} I will now explain how the failure of the unextendability condition (cf.~(a)-(b) above) makes the application of an {\it internal} interpretation problematic, for this particular example.
\\ 
\\
Notice that considering extensions of theories is, according to the perspective of modern QFT, a basic desideratum of any serious theory. The breakdown of Newtonian mechanics at short distances should be seen as an indication of its being an {\it effective theory}.\footnote{Effective field theories 
are theories that are accurate for phenomena in some range of (usually low) energies, but are corrected by higher-order terms in the Hamiltonian, which are relevant at high energies.}

Besides, there is a more specific relation to the gorce-and-morce proposal. Unless the theory is defined to have an exact symmetry, the introduction of new fields will generically introduce higher-order terms that break the seemingly symmetrical way in which those fields appear in the low-energy Hamiltonian. Thus, if gorce and morce are indeed distinct fields,\footnote{The sum of gorce and morce is the derivative of the sum of two potentials. I envisage these two potentials as pertaining to distinct fields---since Hans declares gorce and morce to be distinct.} 
most high-energy theories that reduce to the gorce-and-more theory at low energies, will treat gorce and morce differently. They have different interactions (unless an exact symmetry protects them). Thus the framework of effective field theories promises to satisfy Glymour's demand of parsimony, that there should in principle be a way to determine the values of distinct quantities. 

Thus, the force and the gorce-and-morce theories are {\it generically} not physically equivalent, even though they are Glymour-synonymous. For there are very many possibilities for extension to high energies.
The physical equivalence with the Newtonian force theory can thus only be established if additional requirements are imposed: such as the stipulation of a particular extension of the theory, or a symmetry. This example sheds light on the concrete question of which theories are likely to be unextendable.

\section{Gauge-gravity dualities}\label{ggd}

The remarks in the previous Section, introducing my conceptions of theory and related notions, were necessarily brief. And as I mentioned in \S\ref{concdual}, further work is required to illustrate how my conception of duality, and the contrasts (a) to (d), gives the correct verdicts in cases of dualities that are well-understood (cf.~De Haro and Butterfield (2018)). In this Section, I will use the conception of duality, and the contrasts (a) to (d), to shed light on a less well-understood case, which is of great relevance for our topic, of quantum gravity: gauge-gravity dualities.\footnote{The complete, non-perturbative mathematical theory of gauge-gravity duality is not known. Certain limits of the duality are, however, well-known: the semi-classical limit, in particular. Assuming that some gauge-gravity dualities are exact, at least for a suitable regime of parameters, the scheme illustrates the sense in which two such models are equivalent. For more details, see De Haro et al.~(2017:~\S4.2).}

In \S\ref{sayst}, I take up question (1) in the Introduction: whether spacetime is eliminated; in particular, whether `spacetime' is part of what the theory says. I argue that the common core of the duality includes some, but not all, spacetime structure. In \S\ref{compare}, I discuss other works on duality. In \S\ref{metaph}, I discuss some metaphysical implications.

\subsection{Does what the theories say include `spacetime'?}\label{sayst}

My conception of duality, and the contrasts (a) to (d), give us a way of tackling the question whether `spacetime' is part of the content of the theory, or whether it is eliminated. The conceptions of theories and models in \S\ref{thdual}-\ref{concdual} suggest that, what the theory says---what is physical about it---is the content that the two models share: the content is preserved under the duality map. That content is the triple of states, quantities, and dynamics: viz.~the {\it common core} of the theory. Thus we need to ask: (1) Whether spacetime structures are common to the two models, and thus belong to the common core. (2) Whether there is an internal interpretation which interprets the terms in the same way, in both models. 

Gauge-gravity dualities relate $(d+1)$-dimensional models of quantum gravity to $d$-dimensional quantum field theories (QFT models) with gauge symmetries. And suppose that two such models are dual, in the sense of \S\ref{concdual}. Thus we have an isomorphism between the two Hilbert spaces and between the physical quantities---so we are, in fact, considering unitary equivalence. 
For brevity, I will reduce our two questions, (1) and (2), to the following single question: whether the common core, shared by these two models, is {\it interpreted as spatio-temporal.}

I will next argue that the common core {\it is} spatio-temporal: that it includes a $d$-dimensional spacetime $\cal M$, whose metric is defined only up to local (spacetime-dependent) conformal transformations (De Haro et al.~(2016:~\S\S6.1.2-6.1.3)), i.e.~a conformal manifold. This works as the `core' theory in the following sense. We examine the properties of the two models under the duality map: and the structure which is mapped by duality, once formulated in a model-independent way, is the content of the core triple. In fact, the duality map itself (especially its formulation in De Haro et al.~(2017:~\S4.2)), already makes explicit what the states and the operators of the triple are. So, let us look at the two models and extract their common structure.

To obtain the quantities that the gravity model (under its standard interpretation) takes to be physical, one evaluates the path integral over all metrics and topologies with given {\it boundary conditions}:\footnote{For simplicity, I am now considering the case of quantum gravity without matter. This restricts our considerations to a class of states and operators, with specific conformal dimensions. Adding matter can be done, and does not affect the philosophical conclusions, but would involve some technical qualifications. Also, in the rest of this Section (except for one example with $d=4$), I restrict the discussion to the case where $d$ is odd---again, purely for technical reasons. 
\label{puregrav}} 

(i) The first boundary condition is an asymptotic condition on the form of metric, which is itself determined only up to a conformal factor; in other words, one needs to specify a conformal $d$-dimensional manifold $\cal M$ together with a conformal class of metrics, denoted $[g]$, at the boundary. 

The asymptotic symmetry algebra associated with this model is thus the conformal $d$-dimensional algebra, and the representations of this algebra form the class of possible states that belong to $\cal H$. 

(ii) Second, a boundary condition needs to be imposed on the asymptotic value of the canonical momentum $\Pi_g$ conjugate to the metric induced on the boundary, evaluated on the possible states. 
 This choice further constrains the class of states in ${\cal H}$: it determines a subset of states of the conformal algebra. The simplest choice, $\bra s|\,\Pi_g|s\ket=0$, preserves the full conformal symmetry, and the states $s\in\cal H$ are, accordingly, representations of the $d$-dimensional conformal algebra. Other choices break the conformal symmetry and further constrain the state space.

Thus the boundary condition (ii) fixes a choice of the {\it subset of states} (representations of the conformal algebra) that forms the dynamical Hilbert space of the theory, whereas the first fixes a choice of a {\it source} that is turned on for the canonical momentum $\Pi_g$. We will write the resulting states as $|s\ket_{{\cal M},[g]}\in{\cal H}$, where $s$ is the state, modified by the addition of a source $[g]$ on ${\cal M}$ coupling to $\Pi_g$. The basic physical quantities are the canonical momenta $\Pi_g\in{\cal Q}$ conjugate to $g$ (De Haro et al.~(2017:~\S4.2.2.1)).\footnote{The metric in the interior is not a quantity. There are two ways to see this: (i) from the fact that it is not part of the common core preserved by the duality. But also: (ii) it {\it cannot} be (on the minimalist account of quantities given in this paper), since it is not a diffeomorphism-invariant quantity.}

The quantum field theory model of the theory is a conformal field theory (CFT) on a $d$-dimensional manifold whose metric is defined, up to a local conformal factor, by the very form of the asymptotic metric that one gets from the gravity model: in fact, we can identify this, via duality, with the conformal manifold $\cal M$. The states are representations of the conformal symmetry algebra, the same algebra that we obtain in the gravity model.
The canonical momentum $\Pi_g$ corresponds, through the duality map, to the stress-energy tensor $T_{ij}$ of the CFT: $\Pi_g\equiv T$. The stress-energy tensor is the operator from which the generators of the conformal symmetry algebra can be constructed (see e.g.~Ammon and Erdmenger (2015:~3.2.3)).

Thus, the two models share the $d$-dimensional conformal manifold ${\cal M}$ with its conformal class of metrics $[g]$, the conformal algebra, and the structure of operators, as claimed. The conformal algebra and class determine ${\cal H}$, and thereby the valuations of the important subset $\{T_{ij}\}$ of operators of ${\cal Q}$, i.e.~the infinite set of correlation functions $_{{\cal M},[g]}\bra s|\,T_{i_1j_1}(x_1)\cdots T_{i_nj_n}(x_n)\,|s\ket_{{\cal M},[g]}\,$, where $x_1\,\ldots,x_n\in {\cal M}$, for any $n$. This infinite set of correlation functions contains important dynamical information about the CFT.\footnote{There is no claim here that my description contains {\it all} of the information about the CFT. Nonlocal operators, such as Wilson loops (and perhaps additional states), also need to be compared.} For us, the important point here is that {\it the common core is spatio-temporal.} 


Since my aim is to illustrate the fact that the bare theory contains quantities that can be spatio-temporally interpreted, it will not be necessary to develop the full internal interpretation of the theory. Rather, our next question is whether there is an internal interpretation that supports the spatio-temporal interpretation of the above quantities. 


Two salient elements of an {\it internal interpretation} of the bare theory can be recognised. Notice that {\it both} models interpret the pair $({\cal M},[g])$ and $T_{ij}$ as representing, respectively: (i) a $d$-dimensional conformal manifold with a conformal class of metrics; (ii) stress, energy and momentum. Even if there are other aspects to the internal interpretation of $[g]$ and $T_{ij}$, (i) and (ii) are certainly important elements of it!

We are considering here gravity models with {\it pure gravity}, and no matter fields. The QFT, of course, {\it does} have matter fields. But the stipulation of the {\it specific set} of matter fields is {\it not} part of the invariant core. The interpretation of $T_{ij}$ as the `stress-energy momentum {\it for a specific set of matter fields of the QFT}' can thus not be part of the internal interpretation, because the specific set of matter fields are not part of the common core. Thus, the qualifications of fields as being `gravity', or `matter', are our {\it descriptions} of specific models, and not parts of the theory! 

To illustrate this point, let us discuss the common core and the specific structure further in an example of $d=4$: i.e.~a 4-dimensional QFT and a 5-dimensional gravity model.\footnote{The QFT in question is a `super Yang-Mills theory', a specific supersymmetric variant of Yang-Mills theory, but the details are irrelevant here.}
Specifically, we look at symmetries. The gauge symmetry group of the QFT is $\mbox{SU}(N)$. This symmetry is completely absent from the gravity theory. The QFT is formulated so that this symmetry is explicit---the states ${\cal H}$ and the observables ${\cal Q}$ are invariant under it. But since the common core only contains  states and quantities constructed from the triple $T=\bra{\cal H},{\cal Q},\cal D\ket$, which are invariant under gauge symmetry, this gauge symmetry does {\it not} belong to the common core!

There is thus surprisingly little that is invariant under the duality between these two theories: yet they describe all that is physical about the theories. The $(d+1)$-dimensional manifold $\hat {\cal M}$, most of its diffeomorphism group, gauge symmetries, the dimension of spacetime: in the present context, these are apparently all part of our description, rather than parts of the common core of the theory. Similarly also the concepts of vector fields, tensor fields, Lie groups, differential geometry: though we use them to formulate our theories, each such a concept is not part of nature, at least not part of the common core of the models of a gauge-gravity duality. For example, tensor quantities in $d+1$ dimensions do {\it not} map to tensor quantities in $d$ dimensions under duality, and so do not belong to ${\cal Q}$. Rather, tensors are part of the specific structure!

Thus, the answer to the question we posed in the title of this subsection requires careful articulation, as follows. I state it in terms of what is eliminated from the gravity model:\\
\indent (i) The entire interior region of the $(d+1)$-dimensional manifold, $\hat {\cal M}$ (including its topology), is eliminated.\\
\indent (ii) All that remains is: the {\it asymptotic} conformal manifold $({\cal M},[g])$, which plays the role of asymptotic boundary data for the equations of motion, and so is arbitrary but fixed; and the states $|s\ket_{{\cal M},[g]}\in\cal H$, and the stress-energy tensor $T_{ij}\in{\cal Q}$ at spacelike infinity. This common core is endowed with an action of the conformal group. So, in particular:\\
\indent (iii) All local gravitational structure has been eliminated: there are no {\it local} dynamical gravitational degrees of freedom left. 

In short, `most' of the spacetime structure is eliminated. This agrees with the general expectation in the quantum gravity literature.

\subsection{Comparing with recent work on dualities}\label{compare}

Let us take stock of the distinctions we have made, and discuss how they relate to extant philosophical discussions of dualities in the literature. In this subsection, I will address recent work on dualities and state its limitations, as regards clarifying the {\it contrast between theoretical and physical equivalence}. 

Recent work on dualities engaging with this question includes Matsubara (2013:~p.~485) and Dieks et al.~(2015:~\S3.3.2). And in a special issue on dualities edited by Castellani and Rickles, several authors engage with it. My leading criticism will be that the extant accounts, {\it qua} accounts of dualities, are not sufficiently articulated to provide a clear difference-maker between theoretical and physical equivalence. In defence of these authors, I should add that it was not their aim to look for a clear-cut difference-maker! 

Huggett's (2017) focus seems closest to the ideas developed in this essay. His analysis of T-duality is syntactic, and as such is an interesting alternative to mine, which, as mentioned, is closer to the semantic conception of theory (though I compared with Glymour's syntactic account, in \S\ref{gly}). T-duality, roughly speaking, relates one kind of string theory in a space with a circle of radius $R$ to another kind of string theory in a space with a circle of radius $1/R$. Huggett asks: ``What happens if a duality applies to a `total' theory, in the sense that it is the complete physical description of a world, so that there is nothing outside the theory?'' (p.~86). His position is that, for dual models that describe subsystems, the duals are really distinguishable, by looking outside the model. But theories of the universe lack such a viewpoint, and so no distinction needs to be made between dual models.

We must of course require the bare theories to have correct rules for forming propositions and to be mathematically consistent (\S\ref{intth}, \S\ref{metaph}). But the requirement that the theory be ``a theory of {\it everything}'' could not be taken as a general condition for an internal interpretation: for we do not need the theory to describe absolutely {\it all} the facts, not even all the physical facts, of a world, in order for it to admit an internal interpretation. Rather, what is needed for physical equivalence are the joint requirements of internal interpretation and unextendability (\S\ref{4cond}, \S\ref{unext}).
My construal, in \S\ref{interpretedth}, of a physical interpretation $I_T$ as a pair of partial maps, which can be defined for an unextendable theory on a domain $D_W$, makes this point precise: so that one can now distinguish external and internal interpretations.

One interesting aspect of Rickles' (2017) account is his assertion that theoretical equivalence is a {\it gauge-type} symmetry for {\it all} cases of duality. Our accounts differ in this obvious sense, that mine is stated as an isomorphism between triples, rather that as a gauge-type symmetry.
But the most important difference is that Rickles does not seem to contemplate cases of sophisticated dualities in which the theories could {\it fail to be} physically equivalent. 
If one's account moves too quickly from duality to physical equivalence, it may render the external, and multifaceted, uses of dualities (cf.~the third paragraph of the Introduction, and the last line of \S\ref{unext}) unintelligible. 

Indeed, the account of dualities as just being gauge-type symmetries does not explain the reasons for two theoretically equivalent models' physical inequivalence. The explanation of that {\it is} possible once we develop the idea of an external interpretation.

Finally, I turn to Fraser (2017). My main point will revolve around her chosen example, of Euclidean field theory (EFT) and QFT, not even being a case of theoretical equivalence, let alone physical equivalence. 
Elucidating the distinction between theoretical and physical equivalence seems indeed to be the main aim of Fraser's (2017) example of equivalence between EFT and QFT. She maintains that EFT and QFT are {\it theoretically (formally) equivalent} 
but not physically equivalent. She contrasts this with dualities in string theory, which she describes as cases of {\it physical equivalence}. The example would seem to be well chosen indeed, and Fraser's mastery of QFT is indisputable: but, unfortunately, she does not give a detailed account of how these two cases are supposed to differ.

More importantly, the example {\it itself} is deceptive if taken as a difference-maker for theoretical and physical equivalence: for it is in fact {\it not} a case of theoretical equivalence! As Fraser admits in the first three sections of her essay, EFT and QFT are {\it not} isomorphic: there is a map from EFT to QFT but not the other way around. She writes: ``That the relations are entailments rather than equivalences is not one of the points of comparison between the EFT-QFT case and string theory that I want to emphasize'' (\S3, par.~3). 

One may, of course, choose to downplay the role of equivalence, and focus on a one-way entailment, if one is just interested in a {\it generic} contrast between the EFT-QFT case and dualities in string theory. But one-way entailment will {\it not} give us the difference-maker we need in order to distinguish theoretical from physical equivalence, for the very reason that these are {\it not cases of theoretical equivalence}.\footnote{One might argue that something might still be learned from the contrast between {\it one-way theoretical entailment} and {\it one-way physical entailment}.} As I have argued, the contrast between external and internal interpretations {\it is} such a difference-maker. 

\subsection{What are the broader implications of duality?}\label{metaph}

In this subsection, I discuss the broader implications of gauge-gravity dualities, in three related comments.

The first comment concerns the interpretation of quantum gravity theories. 
In cases in which a duality supports an internal interpretation, duality can be taken to give rise to {\it physical equivalence} (cf.~\S\ref{4cond}). In \S\ref{sayst}, the internal interpretation was developed by stripping the two {\it external interpretations} of irrelevant aspects, and identifying their common core, which constitute (at least parts of) the internal interpretation. This means that the interpretation of a quantum gravity theory, and the articulation of the ontology that may underlie such a theory, starts from the common core of the different models, and the parts of their interpretations that are invariant under duality. This common core can turn out to have surprising properties, e.g.~its spacetime dimension being different from the dimensions of the spacetimes of the models.\footnote{For a more comprehensive discussion of the interpretation of dualities in quantum gravity, see De Haro (2019).}

The second comment concerns physical equivalence. Duality can be found in both unextendable and extendable theories. Gauge-gravity dualities between string or M theory and QFTs are presumed to be exact dualities between unextendable theories, which in particular should have exact formulations.\footnote{For examples in which the duality is known to hold exactly, see De Haro and Butterfield~(2018).} Other dualities, on the other hand, are also exactly defined, but only with limited regimes of applicability (cf.~\S\ref{gly}'s example of the Newtonian force theory and Hans' gorce and morce theory). Such theories are extendable: and for extendable theories we are not always justified in believing internal interpretations, because an extension might modify that interpretation (unless the interpretation is {\it robust}: cf.~footnote \ref{unextendability}). This is because a coherent interpretation is always dependent on how the theory is extended. In such cases, of extendable theories about which we do not know whether they are robust, we are not justified in taking the theories to be physically equivalent. 
The effective field theory perspective in \S\ref{gly} suggested that such theories are generically {\it not} physically equivalent. 


But physical equivalence for {\it unextendable} theories that are theoretically equivalent under an internal interpretation (\S\ref{4cond}) holds good. Such theories
are highly constrained, valid for all values of the parameters, and cannot be coupled to any adjacent physics while preserving their theoretical equivalence. 

This distinction may be seen as a contribution of quantum gravity considerations to discussions of physical equivalence.
It also underlines the importance, for philosophy, of unextendable theories. Extendable theories abound: we have discussed Newtonian mechanics and effective QFTs. There is also general relativity, whose extendability is shown by its singularities due to gravitational collapse. 
Unextendable theories are rare but useful, and we do have some good examples of them: conformal field theories, topological quantum field theories (Chern-Simons theories, various versions of Yang-Mills theory, Wess-Zumino-Witten models) and topological string theories (describing subsectors of string theories). They should provide important {\it case studies for the philosophical topics of theoretical and physical equivalence.}

Thirdly, concerning theory construction: note that unextendable theories need not be `finished' theories. Some theories (especially some two-dimensional conformal field theories) are indeed understood with rigorous mathematics; but other examples (such as Chern-Simons theory and Yang-Mills theory), though expected to be unextendable for good mathematical reasons, are still ``theory fragments'', in the sense of Huggett and W\"uthrich's (2013:~p.~284) quotation in the Introduction.
So, also for the theories mentioned above: completely rigorous mathematical proofs are still lacking, even if the fragments are robust enough that they already contributed to a Fields medal (for E.~Witten: in 1990).

\section*{Envoi}
\addcontentsline{toc}{section}{Envoi}

Let me end by echoing an important remark: the discussion, in Section \ref{ggd}, of the philosophical significance of gauge-gravity dualities, returns us to the idea (echoed, for instance, in the quotation by Huggett and W\"uthrich) that philosophical analysis goes hand-in-hand with theory construction. 

The analysis of gauge-gravity dualities shows specific features of this two-way street. On the one hand, concepts such as {\it theoretical and physical equivalence} (Section \ref{eliminate}) help us construct theories of quantum gravity, and so brings metaphysical analysis to bear on theory construction.

But also, on the other hand: theories of quantum gravity, in particular gauge-gravity dualities, help us achieve greater clarity about those two philosophical concepts, thereby exhibiting the virtues of a `science first' approach to metaphysics.

\section*{Acknowledgements}
\addcontentsline{toc}{section}{Acknowledgements}

I thank Jeremy Butterfield, Nick Huggett, Huw Price, Bryan Roberts, and three anonymous referees for comments on the paper. I also thank several audiences: the British Society for the Philosophy of Science 2016 annual conference, the Oxford philosophy of physics group, LSE's Sigma Club, the Munich Center for Mathematical Philosophy, and DICE2016. This work was supported by the Tarner scholarship in Philosophy of Science and History of Ideas, held at Trinity College, Cambridge. 

\section*{References}
\addcontentsline{toc}{section}{References}

Ammon, M.~and Erdmenger, J.~(2015). {\it Gauge/Gravity Duality. Foundations and Applications}. Cambridge: University Press.\\
\\
Barrett, T.~W.~and Halvorson, H.~(2016). `Glymour and Quine on theoretical equivalence'. {\it Journal of Philosophical Logic,} 45 (5), pp.~467-483.\\
\\
De Haro, S. (2017). `Dualities and emergent gravity: Gauge/gravity duality'.
{\em Studies in History and Philosophy of Modern Physics,} 59, pp.~109-125. \\ 
\\
De Haro, S.~(2019). `The Heuristic Function of Duality'. {\it Synthese,} 196 (12), pp.~5169-5203.\\
\\
De Haro, S.~and Butterfield, J.N.~(2018). `A Schema for Duality, Illustrated by Bosonization'. In:  Kouneiher, J.~(Ed.), {\it Foundations of Mathematics and Physics one century after Hilbert,} pp.~305-376. Springer, Cham. \\ 
\\
De Haro, S., Mayerson, D., Butterfield, J.N. (2016). `Conceptual Aspects of Gauge/Gravity Duality'. {\it Foundations of Physics,} 46, 1381.\\ 
\\
De Haro, S., Teh, N., Butterfield, J.N.~(2017). `Comparing Dualities and Gauge Symmetries'. {\it Studies in History and Philosophy of Modern Physics,} 59, pp.~68-80. \\ 
\\
Dieks, D., Dongen, J. van, Haro, S. de~(2015), `Emergence in Holographic Scenarios for Gravity'. {\it Studies in History and Philosophy of Modern Physics} 52 (B), 203-216.\\ 
\\
Fraser, D.~(2017). `Formal and physical equivalence in two cases in contemporary quantum physics'. {\it Studies in History and Philosophy of Modern Physics,} 59, pp.~30-43. \\
\\
Glymour, C.~(1977). `The epistemology of geometry'. {\it No$\hat{\mbox{u}}$s} 11 (3), 227-251.\\
\\
Huggett, N.~and W\"uthrich, C.~(2013). `Emergent spacetime and empirical (in)coherence'. {\it Studies in History and Philosophy of Modern Physics} 44 (3), 276-285.\\
\\
Huggett, N. (2017), `Target space $\neq$ space'. {\em Studies in the History and Philosophy of Modern Physics,} 59, pp.~81-88. \\
\\
Lewis, D.~(1984). `Putnam's Paradox'. {\it Australasian Journal of Philosophy}, 62 (3), pp.~221-236.\\
\\
Matsubara, K.~(2013). `Realism, underdetermination and string theory dualitites'. {\it Synthese}, 190, pp.~471-489.\\
\\
Rickles, D.~(2012). `AdS/CFT duality and the emergence of spacetime', \emph{Studies in History and Philosophy of Modern Physics}, 44 (3), 312-320.\\
\\
Rickles, D. (2017). `Dual theories: `same but different' or `different but same'?'. {\em Studies in the History and Philosophy of Modern Physics,} 59, pp.~62-67. 

\end{document}